\documentclass[twocolumn,british,prl,superscriptaddress,showpacs]{revtex4-1}
\usepackage[letterpaper]{geometry}
\usepackage[dvips]{graphics}
\usepackage{graphicx}
\usepackage{amsmath}
\usepackage{amssymb}
\usepackage{esint}
\newcommand{\bra}[1]{\langle #1|}
\newcommand{\ket}[1]{|#1\rangle}
\newcommand{\braket}[2]{\langle #1|#2\rangle}
\renewcommand{\t}[1]{\textrm{#1}}

\makeatletter
\@ifundefined{textcolor}{}
{%
 \definecolor{BLACK}{gray}{0}
 \definecolor{WHITE}{gray}{1}
 \definecolor{RED}{rgb}{1,0,0}
 \definecolor{GREEN}{rgb}{0,1,0}
 \definecolor{BLUE}{rgb}{0,0,1}
 \definecolor{CYAN}{cmyk}{1,0,0,0}
 \definecolor{MAGENTA}{cmyk}{0,1,0,0}
 \definecolor{YELLOW}{cmyk}{0,0,1,0}
 }

\makeatother

\makeatother

\usepackage{babel}

\begin{document}

\title{Quantum interferometry with and without an external phase reference}

\author{Marcin Jarzyna, Rafa{\l} Demkowicz-Dobrza{\'n}ski}
\affiliation{Faculty of Physics, University of Warsaw, ul. Ho\.{z}a 69, PL-00-681 Warszawa, Poland}

\begin{abstract}
We discuss the role of an external phase reference in quantum interferometry.
We point out inconsistencies in the literature with regard to the use of the quantum Fisher information (QFI)
in phase estimation interferometric schemes.
We discuss the interferometric schemes with and without an external phase reference
and show a proper way to use QFI in both situations.
\end{abstract}

\pacs{03.65.Ta, 06.20.Dk}

\maketitle

%\begin{figure}
%\includegraphics[width=0.9 \columnwidth]{interferometerclassical.eps}
%\caption{Standard Mach-Zehnder interferometric setup with coherent and squeezed states at the two input ports,
%and photon counting detectors at the outputs.}
%\end{figure}

Laws of quantum mechanics impose fundamental bounds on measurement precisions of basic physical quantities such
as position, momentum, energy, time, phase etc. Theses bounds follow from the structure of the theory itself which contrasts
the situation encountered in classical physics where measurement uncertainties are due to factors which in principle may be
eliminated by improving the quality of measurement procedures. One of the most important measurement techniques
where such bounds have been analyzed is optical interferometry \cite{Hariharan2003}.
% On the practical side, they are the core of world's most elaborated measuring devices such as gravitational wave detectors and atomic clocks.
%In this paper we stay on the theory side and analyze methods based on quantum Fisher information (QFI) that are used to get lower bounds
%imposed by quantum mechanics on phase estimation uncertainties.

%Consider a generic measurement using a Mach-Zehnder interferometer, where a coherent light is
%divided at the input beam-splitter, propagates along two paths of different length accumulating the
%relative phase delay $\varphi$ and is recombined at the output beam-splitter
%making the information on $\varphi$ available through photon-count measurement at the two output ports.
%The shot noise in the detected counts limits the precision of the estimation procedure to
%$\delta \varphi \geq 1/\sqrt{N}$, where $N$ is the average number of photon counts. This bound is referred
%to as the standard quantum limit (SQL), and can be understood at the fundamental quantum level as the result of an
%independent probabilistic behavior of individual photon propagating through the interferometer.

In a generic interferometric measurement using a Mach-Zehnder setup and classical light sources
the precision of estimating the relative phase delay $\varphi$ inside the interferometer is bounded by the
so called standard quantum limit (SQL) $\delta \varphi \geq 1/\sqrt{N}$, where $N$ is average number of photon-counts.
At the fundamental quantum level, the bound is a result of an
independent probabilistic behavior of individual photon propagating through the interferometer.

Breaching the SQL requires the use of special non-classical states of light were photons can no longer be regarded as independent.
One of the first proposals in this direction was the idea to mix coherent light with the squeezed vacuum at the input beam splitter of the Mach-Zehnder interferometer \cite{Caves1981}.
Thanks to the reduced vacuum fluctuations in one of the quadratures of the squeezed state it is
possible to improve the precision beyond the SQL.
This observations prompted the search
for more fundamental bounds on achievable precision, which would be obeyed by all quantum states \cite{Giovannetti2004}.
%, Banaszek2009, Giovannetti2011}. %Three basic elements need to be specified in order to analyze a phase estimation protocol: the input state $\ket{\psi_{\t{in}}}$ that is fed into the interferometer, the measurement $\{\Pi_{n}\}$ that is performed at the output and the estimator $\varphi(n)$ ---
%a function that assigns a phase value to a given measurement outcome.
%In order to find the optimal phase estimation scheme allowed by quantum mechanics one needs to optimize over all input states satisfying certain
%resource constraints (e.g. energy), arbitrary quantum measurements and estimators. This optimization problem can by no means
%be solved by brute-force optimization methods, as the number of free parameters is enormous even for small scale problems.

In general, looking for the optimal phase estimation protocols is difficult since one needs to optimize over
the input state $\ket{\psi_{\t{in}}}$ that is fed into the interferometer, the measurement $\{\Pi_{n}\}$ that is performed at the output and the estimator $\varphi(n)$ --- a function that assigns a phase value to a given measurement outcome.
One of the popular ways to obtain useful bounds in quantum metrology, without the need for cumbersome optimization,
is to use the concept of the quantum Fisher information (QFI) \cite{Helstrom1976}.%, Braunstein1994}.

The purpose of this paper is to give a proper interpretation to the bounds obtained via the QFI
and point out conflicting approaches where seemingly equivalent physical models lead to
different quantitative statements. We show that the source of the problem lies in the use
of quantum states of light which are coherent superpositions of different total photon number terms
without properly taking into account the role of an external phase reference beam.

\begin{figure}[t]
\includegraphics[width=\columnwidth]{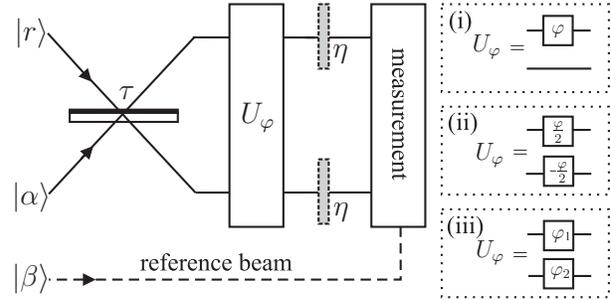}
\caption{An interferometric scheme with coherent and squeezed vacuum states interfered at a beam-splitter,
with arbitrary quantum measurement potentially involving an additional reference beam.
In general, the QFI bounds on the phase estimation precision depend on the way the interferometer phase delay is modeled: (i)
phase shift only in the upper arm, (ii) phase shift distributed
symmetrically, (iii) phase shifts defined with respect to an additional reference beam.
}
\label{fig:interferometer}
\end{figure}
Let $a$, $b$ be the anihilation operators of respectively upper and
lower input modes of the interferometer. For the purpose of this paper we consider the input state of the form
$\ket{\psi_{\t{in}}}= \ket{r,\alpha}$,
where $\ket{\alpha}$ is the coherent state, $a \ket{\alpha} = \alpha \ket{a}$,
while $\ket{r}=\exp[\frac{1}{2}r^* a^2 - \frac{1}{2}r (a^\dagger)^2]\ket{0}$ is the squeezed vacuum state with squeezing parameter $r$
(see Fig.~\ref{fig:interferometer}).
After it has evolved
through the beam-splitter with power transmission $\tau$, and experienced the relative phase shift inside the interferometer $U_\varphi$,
the state becomes $\ket{\psi_\varphi} = U_\varphi B_{\tau} \ket{\psi_{\t{in}}}$
where $B_\tau = \exp[-i\, \t{asin}(\sqrt{\tau})(a^\dagger b + a b^\dagger)]$, $U_\varphi=\exp[-i \varphi a^\dagger a]$.
In a standard Mach-Zehnder setup one
interferes the two modes on another balanced beam-splitter and detects number of photon clicks $n_a$, $n_b$ in
the two output modes. In an idealized setup with no losses, perfect interferometer and 100\% detection efficiency
this leads to a phase dependent probability distribution of clicks:
\begin{equation}
p(n_a,n_b|\varphi)= |\bra{n_a,n_b}B_{1/2} U_\varphi B_{\tau} \ket{\psi_{\t{in}}}|^2.
\end{equation}

Instead of looking for the best possible estimator of the phase, which in general is a hard task, one can
invoke the Cramer-Rao bound \cite{Kay1993} which states that for $k$ repetitions of an experiment and any locally unbiased estimator $\varphi(n_a,n_b)$ the
uncertainty of estimation is bounded from below by:
\begin{equation}
\label{eq:classfish}
\delta \varphi \geq \frac{1}{\sqrt{k F}}, \ F=\sum_{n_a,n_b} \frac{1}{p(n_a,n_b|\varphi)} \left(\frac{d p(n_a,n_b|\varphi)}{d \varphi}\right)^2
\end{equation}
where $F$ is the Fisher information. Moreover, the bound can be saturated in the limit $k \rightarrow \infty$, by making
use of the maximum likelihood estimator.
%$F$ is therefore a useful quantity to assess the amount of information
%that can be extracted from given measurements, without the need to provide an explicit estimator function.
%In our setup, $F$ can be calculated analytically and reads:
%\begin{equation}
%F=
%\end{equation}

A priori it is not obvious that this type of measurement is the optimal way to extract phase information
from the state $\ket{\psi_\varphi}$. The \emph{quantum} Cramer-Rao bound
\cite{Helstrom1976} %, Braunstein1994} 
provides an answer to this problem and states
that whatever the measurement chosen the following bound on the estimation uncertainty holds:
\begin{equation}
\label{eq:fq}
\delta \varphi \geq \frac{1}{\sqrt{k F_Q}}, \
F_Q=4\left(\braket{\psi_\varphi^\prime}{\psi_\varphi^\prime} -
|\braket{\psi_\varphi^\prime}{\psi_\varphi}|^2 \right)
\end{equation}
where  $\ket{\psi_\varphi^\prime}= \frac{d \ket{\psi_\varphi}}{d \varphi}$, and $F_Q$ is called the quantum Fisher information.
$F_Q$ depends neither on the measurement nor on the estimator and
it is solely  a function of the probing state, which makes it an easy to calculate quantity. Moreover, one can always find a
measurement (that may depend on the true value $\varphi$) for which $F=F_Q$.
In what follows, we will drop $k$ for simplicity and use notation where $\delta \varphi \equiv 1/\sqrt{F_Q}$.

A lot of work in quantum enhanced interferometry has been based on utilizing the $F_Q$
\cite{Bollinger1996, Hofmann2009, Ono2010}.
%Huelga1997, Hradil2004, Giovannetti2006, Durkin2007, Fujiwara2008, Dorner2008, Demkowicz2009a, Chwedenczuk2010, Knysh2010, Benatti2010,  Escher2011,
% Genoni2011}.
In a typical approach one maximizes $F_Q$ over a class of input states satisfying some constraint (e.g. total energy)
and in this way finds the input states optimal for quantum interferometry.

Let us investigate the consequences of this
approach in our setup. The input state $\ket{\psi_{\t{in}}}=\ket{\alpha,r}$ has the mean number of photons equal to:
$\bar{n}= |\alpha|^2 + \sinh^2 r$. Having fixed $\bar{n}$ we look for optimal $\alpha, r$
and the transmission coefficient $\tau$ that maximize $F_Q$.
If one follows this procedure rigorously, then
%one arrives at a physically counterintuitive result that
the solution
depends strongly on the way the phase shift between the beams is modeled inside the interferometer.
 As a simple illustration of this counterintuitive behavior,
the relative phase shift $\varphi$ may be modeled in e.g. two ways depicted in Fig.~\ref{fig:interferometer} as (i) and (ii).
These two cases correspond $U^{(i)}_\varphi =\exp[-i \varphi a^\dagger a]$ and $U^{(ii)}_\varphi=\exp[- i \frac{\varphi}{2} a^\dagger a
+ i \frac{\varphi}{2} b^\dagger b]$ respectively. When plugged into Eq.~(\ref{eq:fq}) they yield:
\begin{eqnarray}
F_Q^{(i)} &= 4 \tau^2 |\alpha|^2 +2(1-\tau)^2 \sinh^2(2r)+ \mathfrak{F}
\label{eq:fqi}
\\
F_Q^{(ii)} &= (1-2\tau)^2\left[ |\alpha|^2 +\frac{1}{2} \sinh^2(2r)\right] +\mathfrak{F}
\label{eq:fqii}
\end{eqnarray}
where $\mathfrak{F}=4 \tau (1-\tau)\left(|\alpha|^2 e^{2r}+\sinh^2r\right)$, and
in order to simplify the formulas we have put the relative phase between the input beams to be $\pi/2$ ($r = |r|$, $\alpha=i |\alpha|$),
which is the optimal choice for this and all the examples presented in this paper.
%$\mathfrak{F}=4 \tau (1-\tau)\left[|\alpha|^2 (1+2\sinh^2 r - \sinh 2r \cos 2\theta)+\sinh^2r\right]$, where $\theta$ is the
%relative phase between the input beams. For simplicity, in what follows we put  $r=|r|$,
%$\alpha=|\alpha|e^{i \theta}$.
The formulas are clearly different,
which becomes evident when we set $\tau=1/2$, $r=0$, in which case $F_Q^{(i)}=2|\alpha|^2$, $F_Q^{(ii)}=|\alpha|^2$.

To understand what lays behind this discrepancy consider an even more exotic case of $\tau=1$, $r=0$.
The coherent state is simply transmitted to the upper arm
so there is no interferometer at all, yet
$F_Q^{(i)}=4 |\alpha|^2$, $F_Q^{(ii)}=|\alpha|^2$.
To give a meaning to these ,,unphysical'' results, notice that QFI simply depends on the change of the probe state
under the variation of the parameter $\varphi$. Even if we send a coherent state $\ket{\alpha}$ to the upper arm alone,
then under the phase shift $\varphi$ it evolves to $\ket{\alpha e^{i\varphi}}$, which differs from $\ket{\alpha}$ and
in principle may provide us with useful information on the value of phase $\varphi$. The physical content that is missing in this reasoning
is that the phase information is only available once we have access to an additional reference beam with respect to which the phase shift $\varphi$
is defined. In other words there is no such thing as an absolute phase shift --- a seemingly obvious fact
which has nevertheless significant implications for the problem considered, and has been
treated in contradicting ways in the literature.
%is not always properly addressed in the literature on the subject.

The whole problem revolves around quantum states that are coherent superpositions of different
photon number states such as e.g. coherent or squeezed states. Take a coherent state $\ket{|\alpha|e^{i \theta}}$.
Since all measurements in quantum optics
rely ultimately on photon counts, no measurable consequences of these coherences may be observed
unless these states are interfered (as e.g. in a homodyne measurement)
with a reference beam with respect to which the phase $\theta$ is defined.
Otherwise,
%if the reference beam is not available,
one is entitled to phase average the state
without any observable consequences, i.e. replace $\ket{|\alpha|e^{i \theta}}$ with
$\rho= \int \frac{d\theta}{2\pi} \ket{|\alpha|e^{i \theta}}\bra{|\alpha|e^{i \theta}}$ which is an incoherent
mixture of photon number states with Poissonian statistics \cite{Molmer1997}.%, Bartlett2006, Hyllus2010}.

Going back to our quantum interferometric setup, if we indeed consider just the two modes of the interferometer and do not allow
any additional reference beam, then as an input we should rather consider a phase averaged state of the form:
\begin{equation}\label{eq:ro}
\rho(r, \alpha) = \int \frac{d \theta}{2 \pi} V^a_\theta V^b_\theta \ket{r,\alpha}\bra{r,\alpha} V^{a \dagger}_\theta V^{b \dagger}_\theta
\end{equation}
where $V_\theta^{x} =\exp(- i \theta x^\dagger x)$. Notice, that squeezed and coherent states are averaged over a common phase $\theta$,
which reflects the fact there is a physical meaning in the \emph{relative} phases between them.
Calculation of QFI for $\rho(r,\alpha)$ --- $F_Q^{(\rho)}$--- are more involved since the state is mixed
and instead of Eq.~(\ref{eq:fq}) one needs to employ a general formula involving the concept of the symmetric logarithmic derivative \cite{Helstrom1976}.
The resulting $F_Q^{(\rho)}$ is different both from $F_Q^{(i)}$ and $F_Q^{(ii)}$ and
does not depend on the choice of the phase shift generator---
be it $U_\varphi^{(i)}$ or $U^{(ii)}_\varphi$. All that matters is the relative phase between
the arms of the interferometer.
$F_Q^{(\rho)}$ achieves maximum for $\tau=1/2$ in which case it takes a simple form:
%Final result is a bit complicated, but without going into details, it can be shown, that Fischer information is maximal for $\tau=\frac{1}{2}$ and %equal:
\begin{equation}
\label{eq:fqrho}
\max_\tau F_Q^{(\rho)}= F_{Q,\tau=1/2}^{(\rho)}  = |\alpha|^2 e^{2r}+\sinh^2r.
\end{equation}

There is a great deal of confusion in the literature since formulas $F_Q^{(i)}$, $F_Q^{(ii)}$ are often used
instead of $F_Q^{(\rho)}$ without discussing the need of an additional reference beam \cite{Ono2010, Joo2011, Spagnolo2011}. Despite this, one sometimes
arrives at the correct result, since e.g. for $\tau=1/2$, $F_Q^{(ii)}=F_Q^{(\rho)}$ and that is why the results in
 \cite{Ono2010} are indeed correct. However, had one used a phase shift generator (ii) instead of (i) one would arrive at a different solution.
Similar objections can be raised in the context of \cite{Joo2011} where $F_Q^{(i)}$ is used and \cite{Spagnolo2011} where
one defines standard quantum limit as $\delta \varphi^2 = (2|\alpha|^2)^{-1}$ instead of $(|\alpha|^2)^{-1}$
which is again due to the use of $F_Q^{(i)}$ instead of $F_Q^{(\rho)}$.
Making use of $F_Q^{(i)}$, $F_Q^{(ii)}$ without mentioning the need of a reference beam is misleading since
it is not clear what experimental setup these quantities really refer to.
%especially when one quantifies the resources needed in the experiment. The additional reference beam clearly contributes energy
%and if the total energy is the resource that one is concerned with one should take this into account.

Let us now consider a situation in which we indeed have an access to an additional reference beam --- represented by the state $\ket{\beta}$ in Fig.\ref{fig:interferometer} ---
 and want to properly analyze the quantum interferometric setup. If the reference beam is strong, we can treat it as a phase reference
 for the other two modes. Therefore we introduce two phase shifts $\varphi_1$, $\varphi_2$ as in (iii) in Fig.~\ref{fig:interferometer}
 which are defined with respect to the reference beam. In a sense, we now face a two-parameter estimation problem.
 The proper way to proceed is to employ a two parameter Cramer-Rao bound \cite{Helstrom1976}:
 \begin{equation}
 \Sigma \geq \mathcal{F}^{-1}, \ \mathcal{F}_{ij}=4 \Re \left(\braket{\partial_i \psi}{\partial_j \psi}
 -  \braket{\partial_i \psi}{\psi} \braket{\psi}{\partial_j \psi}\right),
 \end{equation}
 where $\Sigma_{ij}$, $i=1,2$ is the covariance matrix for parameters $\varphi_1$, $\varphi_2$,
 $\mathcal{F}$ is the quantum Fisher information matrix (QFIM),
 $\ket{\psi} = V^a_{\varphi_1} V^b_{\varphi_2} B_{\tau} \ket{\psi_{\t{in}}}$ is the probe state after sensing the phase shifts
 $\varphi_1$, $\varphi_2$ and $\ket{\partial_i \psi}=\frac{\partial \ket{\psi}}{\partial \varphi_i}$.
 If one is now interested in the bound on the uncertainty of $\varphi_i$ the proper formula reads:
 \begin{equation}\label{eq:prec1}
 \delta \varphi_i \geq \sqrt{(\mathcal{F}^{-1})_{ii}}.
 \end{equation}
Note that in general $(\mathcal{F}^{-1})_{ii} \neq (\mathcal{F}_{ii})^{-1}$.

In quantum interferometry we are interested in the phase shift difference between the interferometer arms i.e.
$\varphi_{-} = \varphi_1 - \varphi_2$, so it is more convenient to write QFIM in basis
$\varphi_{\pm}= \varphi_1 \pm \varphi_2$. Calculating QFIM in $\pm$ basis yields:
\begin{equation}
\mathcal{F}=
\left(\begin{array}{cc}
\mathfrak{G} & (1-2\tau) \mathfrak{H} \\
(1-2\tau) \mathfrak{H}& (1-2\tau)^2 \mathfrak{G} + \mathfrak{F}
\end{array} \right)
\end{equation}
%\begin{align}
%\mathcal{F}_{++} &= \mathfrak{G},
%\quad \mathcal{F}_{--} = (1-2\tau)^2 \mathfrak{G} + \mathfrak{F}, \\
%F_{+-}&  =F{-+} = (1-2\tau) \mathfrak{G}
%\end{align}
where $\mathfrak{G} = |\alpha|^2 +\sinh^2(2r)/2$, $\mathfrak{H} = \sinh^2(2r)/2-|\alpha|^2$, and finally  the bound on estimation precision of $\varphi_-$ can obtained easily via Eq.~(\ref{eq:prec1}). The minimal uncertainty is obtained for $\tau=\frac{1}{2}$
in which case $\delta\varphi_-\geq\sqrt{\left(|\alpha|^2e^{2r} +\sinh^2r\right)^{-1}}$.
%\begin{equation}
%\label{eq:prbound}
%\delta \varphi_- \geq \sqrt{(\mathcal{F}^{-1})_{--}}.
%\end{equation}
%This result is a bit complicated, however, for $\tau=\frac{1}{2}$, it simplifies to $\delta\varphi_-\geq\sqrt{\mathfrak{F}}$.

It is interesting to note that this is the same result as the one obtained for the phase averaged state using $F_Q^{(\rho)}$ from
Eq.~(\ref{eq:fqrho}). %, which is a more general property of pure path-symmetric states \cite{Jarzyna2011}.
%Since
%$\mathcal{F}$ was calculated under the assumption of availability of an additional reference beam,
%$F_Q^{(\rho)}$ was calculated assuming no additional reference beam,
This observation proves that in the setup considered (with $\tau=\frac{1}{2}$)
there is no advantage in using the reference beam when estimating the phase difference between the two arms of the interferometer.
More generally, it can be shown that this is a feature of all path-symmetric pure states, i.e the states that are symmetric with respect to an exchange of the arms of the interferometer \cite{Jarzyna2011}. It is also worth mentioning that the optimal measurement in our setup
when $\tau=1/2$, and more generally whenever we deal with a pure path-symmetric state in the interferometer, is a standard
photon count measurement after the two modes are interfered on a balanced beam splitter \cite{Pezze2008, Hofmann2009}.

\begin{figure}
\includegraphics[width=\columnwidth]{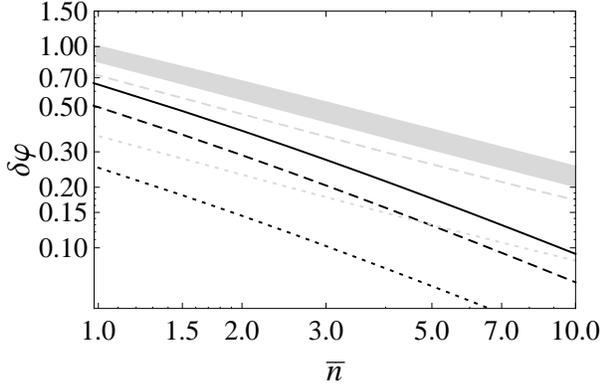}
\caption{Bounds on the phase estimation precision calculated using the QFI, in case of an ideal (black)
and lossy ($\eta=0.8$, gray) interferometer. Different curves correspond to QFI calculated using different models:
$F_Q^{(i)}$ (dotted), $F_Q^{(ii)}$ (dashed), $F_Q^{(\rho)}$ (solid). In case of a lossy interferometer
the additional reference beam may improve the precision:
$(\mathcal{F}^{-1})_{--} \t{ (gray, solid, bottom) } < \left(F_Q^{(\rho)}\right)^{-1} \t{ (gray, solid, top) }$,
while for the ideal interferometer these quantities coincide.
}
\label{fig:compnoloss}
\end{figure}

As a summary of the discussion, in Fig.~\ref{fig:compnoloss} we plot in black the bounds on $\delta \varphi$ obtained using
different QFIs. The bounds are plotted as a function of the total number of photons $\bar{n}$ used,
and parameters ($\tau$, $\alpha$, $r$) are chosen to maximize the respective QFI.
One can easily notice that the uncertainties calculated using $F_Q^{(i)}$ and $F_Q^{(ii)}$
are overly optimistic. The reason behind this is an implicit assumption that e.g. in the case of $F_Q^{(i)}$,
the lower arm of the interferometer (where there is no phase shift element)
is perfectly aligned with the reference beam. Such an assumption can hardly be justified in practice.
%\begin{figure}
%\includegraphics[width=\columnwidth]{comparisonloss.eps}
%\caption{Comparison of bounds on phase estimation precision using different approaches for $\eta=0.8$.}
%\label{fig:comploss}
%\end{figure}

Things become more complicated when one takes into account loss in the interferometer.
%We model loss by fictitious beam-splitters with power transmission
Let $\eta$ be the power transmission coefficient in both arms of the interferometer.
All results presented in the paper may be rederived in this setup although calculations are more involved.
Fig.~\ref{fig:compnoloss} depicts in gray the resulting uncertainties for exemplary loss coefficient
 $1-\eta=0.2$. Apart from a similar observation that $F_Q^{(i)}$ and $F_Q^{(ii)}$
yield overoptimistic results, we additionally observe that $(\mathcal{F}^{-1})_{--} < \left(F_Q^{(\rho)}\right)^{-1}$
which is illustrated by a thick band and proves that having an additional reference beam helps in
estimating the phase difference in a lossy interferometer.

It is interesting to understand deeper what we really mean by \emph{strong} reference beam. Clearly if
$|\beta|$ is not strong enough we can hardly treat it as a phase reference.
To solve this problem consider a phase averaged three mode state:
\begin{equation}
\rho(r,\alpha,\beta) = \int \frac{d \theta}{2 \pi} V^a_\theta V^b_\theta V^c_\theta \ket{r, \alpha,\beta}
\bra{r, \alpha,\beta} V^{a \dagger}_\theta V^{b \dagger}_\theta V^{c \dagger}_\theta.
\end{equation}
%which is a proper way to proceed if we do not have access to an additional (fourth... ufff)
%reference beam.
Calculating the QFIM in this case can be done only numerically. Finally we can calculate
the optimal estimation strategy (optimal $\tau$, $\alpha$, $r$) and the resulting
bound on precision $\delta\varphi_- \geq (\mathcal{F}^{-1})_{--}$ as a function of $|\beta|$.
%The inset of Fig.~\ref{fig:comploss}  depicts the improvement of optimal estimation precision with
%ncreasing strength of the reference beam, for $\bar{n}=0.5$.
With the increasing value of $|\beta|$ we will approach the regime discussed before
where we treated the reference beam as strong enough so it can serve as a perfect phase reference.
In the case of an example depicted in gray in Fig.~\ref{fig:compnoloss} this
corresponds to improving the estimation precision by going form the upper to the lower boundary of the gray band with increasing $|\beta|$.

A deeper analysis \cite{Jarzyna2011} shows that a sufficient condition for treating the reference beam as a perfect phase reference
is $|\beta|^2 \gg \bar{n}^2$.
The reference beam
needs to have much more than the square of the number of photons traveling in the proper modes of the interferometer,
a fact observed also in \cite{Anisimov2011}.

In summary we have pointed out some possible flaws in the interpretations of the results obtained using the QFI for
states which are superpositions of different total photon number terms and showed that the full understanding of the problem is only possible
if the role of an additional reference beam is properly taken into account.

We acknowledge interesting exchange of ideas with Petr Anisimov and Jonathan Dowling.
We thank Konrad Banaszek for initiating the project and many inspiring discussions. This research was supported by the European Commission under
the Integrating Project Q-ESSENCE and the Foundation for Polish Science under the TEAM programme.

\end{document}